\documentclass[prl,twocolumn,10pt,aps]{revtex4-1}

 \usepackage{bm}
 \usepackage{verbatim}
 \usepackage{amsmath}
 \usepackage{amssymb}
 \usepackage{amsthm}
 \usepackage{latexsym}
 \usepackage{amsfonts}
 \usepackage{epsfig}
 \usepackage{epstopdf}
 \usepackage{wasysym} % for certain special symbols
 \usepackage{subfigure}
 \usepackage{color}
 \definecolor{darkblue}{rgb}{0,0,.5}
 \usepackage[linktocpage, colorlinks=true, linkcolor=darkblue, citecolor=darkblue]{hyperref}
 \usepackage[all]{hypcap}
 \usepackage[makeroom]{cancel}

\newcommand{\bb}[1]{\textbf{#1}}

\begin{document}

\title{Reply to comment on ``Measurability of nonequilibrium thermodynamics in terms of the Hamiltonian of mean force''}

\author{Philipp Strasberg$^1$}
\author{Massimiliano Esposito$^2$}
%\author{Mar\'ia Garc\'ia D\'iaz}
%\author{Andreas Winter$^{1,2}$}
%\email{philipp.strasberg@uab.cat}
\affiliation{$^1$F\'isica Te\`orica: Informaci\'o i Fen\`omens Qu\`antics, Departament de F\'isica, Universitat Aut\`onoma de Barcelona, 08193 Bellaterra (Barcelona), Spain}
\affiliation{$^2$Complex Systems and Statistical Mechanics, Departement of Physics and Materials Science, University of Luxembourg, L-1511 Luxembourg, Luxembourg}
%\affiliation{$^2$ICREA -- Instituci\'o Catalana de Recerca i Estudis Avan\c{c}ats, Pg.~Lluis Companys, 23, 08010 Barcelona, Spain}

\date{\today}

\begin{abstract}
We refute the criticism expressed in a comment by P. Talkner and P. H\"anggi [Phys. Rev. E \bb{102}, 066101 (2020)] 
on our paper Phys. Rev. E \bb{101}, 050101(R) (2020). We first make clear that our paper is free of any technical 
mistakes. We then defend the statements of our manuscript which were claimed to be taken out of context. 
Finally, we give clear arguments showing that the basic concepts we rely on are meaningful and have a deep rational. 
\end{abstract}

\maketitle
%\tableofcontents

%\newtheorem{mydef}{Definition}[section]
\newtheorem{lemma}{Lemma}[section]

In our recent paper~\cite{StrasbergEspositoPRE2020}, we were concerned with setting up a nonequilibrium thermodynamic 
framework for a system in strong contact with a bath using the concept of the Hamiltonian of mean force (HMF). 
A recent comment by Talkner and H\"anggi (abbreviated T\&H in the following) claims that our results can 
``lead to erroneous conclusions'' and that we ``wrongly conclude'' some things~\cite{TalknerHaenggiPRE2020}. 

The first statement was made in the context of a system-bath Hamiltonian of the form~\cite{TalknerHaenggiPRE2020} 
\begin{equation}\label{eq Hamiltonian SB all}
 H_\text{tot}(\mu) = H_S(\mu) + H_{SB}(\mu) + H_B(\mu).
\end{equation}
Here, $\mu$ denotes all externally controllable parameters (such as magnetic or electric fields) appearing either 
in the system Hamiltonian $H_S(\mu)$, the bath Hamiltonian $H_B(\mu)$ or their mutual interaction term $H_{SB}(\mu)$. 
The setup~(\ref{eq Hamiltonian SB all}) is \emph{not} the one that was considered in our 
paper~\cite{StrasbergEspositoPRE2020}. We explicitly stated that the external driving does not act on the 
bath and the system-bath interaction. Concluding in the more general setup of 
Eq.~(\ref{eq Hamiltonian SB all})~\cite{TalknerHaenggiPRE2020} that 
information about the system is not sufficient to characterize the nonequilibrium thermodynamics is obvious, and was 
clearly not the statement made in our paper. Thus, within the stated assumptions of our 
paper, no evidence is provided against the consistency and correctness of our results. 
We also remark that Eq.~(\ref{eq Hamiltonian SB all}) was also not the starting point 
used by T\&H in their recent papers~\footnote{For instance, in Ref.~\cite{TalknerHaenggiPRE2016}.
T\&H state that the ``time-dependent parameters $\lambda$ [...] only enter the system Hamiltonian and are supposed to 
be externally controllable'' and ``the interaction with the environment as well as the Hamiltonian of the bare 
environment are considered as being independent of $\lambda$'' (note that $\mu$ was denoted $\lambda$ therein); also 
see Eq.~(17) in Ref.~\cite{TalknerHaenggiArXiv2019}. We remark that we refer to version 2 on the arXiv of 
Ref.~\cite{TalknerHaenggiArXiv2019}, which was the version available to us at the time of writing our 
paper~\cite{StrasbergEspositoPRE2020}.} to which we refer in Ref.~\cite{StrasbergEspositoPRE2020}. 

Next, we come to the statement that we ``wrongly conclude''~\cite{TalknerHaenggiPRE2020} something in footnote~[60] 
of Ref.~\cite{StrasbergEspositoPRE2020}. For that purpose, we denote by $p(x,t)$ the classical phase space probability 
distribution of the open system, where $x$ denotes the system phase space coordinates. Now, 
in~\cite{TalknerHaenggiPRE2016} T\&H argue that the definition of thermodynamic quantities suffers from ambiguities 
since it is possible to add any function $h(x)$ of the system's phase space coordinates to them as long as 
\begin{equation}\label{eq ambiguities}
 \int dx h(x)p(x,t) = 0
\end{equation}
holds. Since Eq.~(\ref{eq ambiguities}) is supposed to hold for all distributions $p(x,t)$, which includes in 
particular a basis of functions $p(x,t)$ (regarded as elements of a Hilbert space), our footnote~[60] only states 
that this implies $h(x) \equiv 0$, which is a mathematical fact. 
Note that this argument continues to hold if $h(x) = h(x,t)$ is an arbitrary prescribed function of time. 
To safe the argument of Ref.~\cite{TalknerHaenggiPRE2016}, T\&H now claim that $h$ also ``changes with time in 
dependence of the initial probability density function of the system''~\cite{TalknerHaenggiPRE2020}. However, even if 
we always start with the same initial state $p(x,0)$, we can still generate different final states $p(x,t)$ ($t>0$) 
by using a set of different driving protocols $\{\mu_s|0\le s\le t\}$ while keeping the final value $\mu_t$ fixed. 
Equation~(\ref{eq ambiguities}) then has to hold with respect to this set of final states, which still 
severely restricts the admissible $h$. Only if 
\begin{equation}\label{eq h}
 h(x) = h[x,\{\mu_s|0\le s\le t\},p(x,0)]
\end{equation}
is a functional of the initial system state \emph{and} the entire driving protocol, our argument looses its validity. 
However, identifying ambiguities that require a precise and complicated fine-tuning in every run of the experiment 
seems questionable, and we also can not find the ansatz~(\ref{eq h}) in the original writing of 
T\&H~\cite{TalknerHaenggiPRE2016}. 

We now turn to the claim that our paper contains ``two misleading literal citations''~\cite{TalknerHaenggiPRE2020}. 
The first claim refers to the quote ``...presents in practice an impossible task'' taken from 
Ref.~\cite{TalknerHaenggiArXiv2019}, which---in view of T\&H---``does not refer to the HMF [...] but to the 
reconstruction of the total [...] Hamiltonian''~\cite{TalknerHaenggiPRE2020}. What is true is that the subject of 
the sentence we quote is the total Hamiltonian and not the HMF. What is also true, however, is that knowledge of the 
total Hamiltonian allows to compute (in principle) the HMF and that a preceeding sentence in 
Ref.~\cite{TalknerHaenggiArXiv2019} reads ``Therefore, the Hamiltonian of mean force cannot be obtained from a purely 
system-intrinsic point of view [...].'' Hence, our quotation is \emph{not} ``taken out of [its] original 
context''~\cite{TalknerHaenggiPRE2020}. 
The second ``misleading literal citation''~\cite{TalknerHaenggiPRE2020} refers to the fact that we claim that 
the assessment of work in the quantum regime by T\&H is based on measurements that ``need to be 
error free''~\cite{StrasbergEspositoPRE2020}. When looking at Ref.~\cite{TalknerHaenggiArXiv2019}, we can 
only find that T\&H introduce work in the quantum regime via the two-point measurement scheme~\footnote{Towards the end 
of Sec.~IV.A.2 of Ref.~\cite{TalknerHaenggiArXiv2019}, T\&H also cite some literature using generalized measurements, 
but they still conclude that ``the measurements yet need to be error free''~\cite{TalknerHaenggiArXiv2019}.}. It is a 
fact that this scheme, also in the way it is reviewed in Ref.~\cite{TalknerHaenggiArXiv2019}, requires perfect 
error-free measurements. Therefore, we must conclude that any comment concerning work in the quantum regime made in 
Ref.~\cite{TalknerHaenggiArXiv2019} refers to this particular measurement scheme. 

Finally, we turn to two general claims of T\&H. First, it is claimed that the use of the HMF outside equilibrium 
considerations is ``without a deeper rational''~\cite{TalknerHaenggiPRE2020}. We believe that the 70 references given 
in our paper can be seen as convincing counterevidence. In fact, the HMF was successfully used in even more 
nonequilibrium contexts without explicitly noting it (which might be related to the 
fact that this terminology is non-standard in physics). Also, as we have demonstrated in an earlier 
paper~\cite{StrasbergEspositoPRE2017}, the HMF framework emerges naturally whenever a system is strongly coupled to 
a \emph{fast} environment. We started our investigation in Ref.~\cite{StrasbergEspositoPRE2017} based on a Markov 
process obeying local detailed balance, but the argument can be generalized by using projection operator 
techniques. 

Second, T\&H claim that the fluctuating probability density function used in stochastic thermodynamics, e.g., to 
compute a stochastic entropy ``does not stay normalized in general'' (supplement of Ref.~\cite{TalknerHaenggiPRE2020}) 
and ``has no obvious probabilitic meaning''~\cite{TalknerHaenggiPRE2020}. We disagree. The construction of this 
fluctuating probability density function is well defined, although it experimentally requires \emph{pre}processing of 
many recorded stochastic trajectories. This makes its application not straighforward, but the theory of classical 
stochastic processes---by virtue of the Kolmogorov consistency condition~\cite{KolmogorovBook2018}---guarantees 
that this probability density function is normalized and non-negative. Thus, also any functional defined it terms of 
it and evaluated along stochastic trajectories gives rise to a well defined stochastic process. 

\emph{In conclusion,} we made clear that our paper~\cite{StrasbergEspositoPRE2020} does not contain technically or 
contextually wrong statements. Moreover, we feel that a significant part of the criticism~\cite{TalknerHaenggiPRE2020} 
was not targeted at our paper, but also applies to weak coupling thermodynamics or, more generally, any attempt to 
formulate thermodynamics out of equilibrium. We clearly refute this criticism (which has also been already debated in 
other places): the foundations of stochastic thermodynamics are sound. Finally, we acknowledge that many open 
questions remain outside the traditional scope of stochastic thermodynamics, but we remain confident that they can be 
also addressed in a constructive way.

%%%%%%%%%%%%%%%%%%%%%%%%%%%%%%%%%%%%%%%%%%%%%%%%%%%%%%%%%%%%%%%%%%%%%%%%%%%%%%%%%%%%%%%%%%%%%%%%%%%%%%%%%%%%%%%%%%%%%%%%
\emph{Acknowledgements.---}PS is financially supported by the DFG (project STR 1505/2-1) and also acknowledges 
funding from the Spanish MINECO FIS2016-80681-P (AEI-FEDER, UE) and FEDER/Ministerio de Ciencia e Innovaci\'on -- 
Agencia Estatal de Investigaci\'on, project PID2019-107609GB-I00. ME is supported by the European Research Council 
project NanoThermo (ERC-2015-CoG Agreement No. 681456). 

%%%%%%%%%%%%%%%%%%%%%%%%%%%%%%%%%%%%%%%%%%%%%%%%%%%%%%%%%%%%%%%%%%%%%%%%%%%%%%%%%%%%%%%%%%%%%%%%%%%%%%%%%%%%%%%%%%%%%%%%

\bibliography{/home/philipp/Documents/references/books,/home/philipp/Documents/references/open_systems,/home/philipp/Documents/references/thermo,/home/philipp/Documents/references/info_thermo,/home/philipp/Documents/references/general_QM,/home/philipp/Documents/references/math_phys,/home/philipp/Documents/references/equilibration}

\begin{thebibliography}{8}%
\makeatletter
\providecommand \@ifxundefined [1]{%
 \@ifx{#1\undefined}
}%
\providecommand \@ifnum [1]{%
 \ifnum #1\expandafter \@firstoftwo
 \else \expandafter \@secondoftwo
 \fi
}%
\providecommand \@ifx [1]{%
 \ifx #1\expandafter \@firstoftwo
 \else \expandafter \@secondoftwo
 \fi
}%
\providecommand \natexlab [1]{#1}%
\providecommand \enquote  [1]{``#1''}%
\providecommand \bibnamefont  [1]{#1}%
\providecommand \bibfnamefont [1]{#1}%
\providecommand \citenamefont [1]{#1}%
\providecommand \href@noop [0]{\@secondoftwo}%
\providecommand \href [0]{\begingroup \@sanitize@url \@href}%
\providecommand \@href[1]{\@@startlink{#1}\@@href}%
\providecommand \@@href[1]{\endgroup#1\@@endlink}%
\providecommand \@sanitize@url [0]{\catcode `\\12\catcode `\$12\catcode
  `\&12\catcode `\#12\catcode `\^12\catcode `\_12\catcode `\%12\relax}%
\providecommand \@@startlink[1]{}%
\providecommand \@@endlink[0]{}%
\providecommand \url  [0]{\begingroup\@sanitize@url \@url }%
\providecommand \@url [1]{\endgroup\@href {#1}{\urlprefix }}%
\providecommand \urlprefix  [0]{URL }%
\providecommand \Eprint [0]{\href }%
\providecommand \doibase [0]{http://dx.doi.org/}%
\providecommand \selectlanguage [0]{\@gobble}%
\providecommand \bibinfo  [0]{\@secondoftwo}%
\providecommand \bibfield  [0]{\@secondoftwo}%
\providecommand \translation [1]{[#1]}%
\providecommand \BibitemOpen [0]{}%
\providecommand \bibitemStop [0]{}%
\providecommand \bibitemNoStop [0]{.\EOS\space}%
\providecommand \EOS [0]{\spacefactor3000\relax}%
\providecommand \BibitemShut  [1]{\csname bibitem#1\endcsname}%
\let\auto@bib@innerbib\@empty
%</preamble>
\bibitem [{\citenamefont {Strasberg}\ and\ \citenamefont
  {Esposito}(2020)}]{StrasbergEspositoPRE2020}%
  \BibitemOpen
  \bibfield  {author} {\bibinfo {author} {\bibfnamefont {P.}~\bibnamefont
  {Strasberg}}\ and\ \bibinfo {author} {\bibfnamefont {M.}~\bibnamefont
  {Esposito}},\ }\href {\doibase 10.1103/PhysRevE.101.050101} {\bibfield
  {journal} {\bibinfo  {journal} {Phys. Rev. E}\ }\textbf {\bibinfo {volume}
  {101}},\ \bibinfo {pages} {050101} (\bibinfo {year} {2020})}\BibitemShut
  {NoStop}%
\bibitem [{\citenamefont {Talkner}\ and\ \citenamefont
  {H\"anggi}(2020)}]{TalknerHaenggiPRE2020}%
  \BibitemOpen
  \bibfield  {author} {\bibinfo {author} {\bibfnamefont {P.}~\bibnamefont
  {Talkner}}\ and\ \bibinfo {author} {\bibfnamefont {P.}~\bibnamefont
  {H\"anggi}},\ }\href {\doibase 10.1103/PhysRevE.102.066101} {\bibfield
  {journal} {\bibinfo  {journal} {Phys. Rev. E}\ }\textbf {\bibinfo {volume}
  {102}},\ \bibinfo {pages} {066101} (\bibinfo {year} {2020})}\BibitemShut
  {NoStop}%
\bibitem [{Note1()}]{Note1}%
  \BibitemOpen
  \bibinfo {note} {For instance, in Ref.~\cite {TalknerHaenggiPRE2016}. T\&H
  state that the ``time-dependent parameters $\lambda $ [...] only enter the
  system Hamiltonian and are supposed to be externally controllable'' and ``the
  interaction with the environment as well as the Hamiltonian of the bare
  environment are considered as being independent of $\lambda $'' (note that
  $\mu $ was denoted $\lambda $ therein); also see Eq.~(17) in Ref.~\cite
  {TalknerHaenggiArXiv2019}. We remark that we refer to version 2 on the arXiv
  of Ref.~\cite {TalknerHaenggiArXiv2019}, which was the version available to
  us at the time of writing our paper~\cite
  {StrasbergEspositoPRE2020}.}\BibitemShut {Stop}%
\bibitem [{\citenamefont {Talkner}\ and\ \citenamefont
  {H\"anggi}(2016)}]{TalknerHaenggiPRE2016}%
  \BibitemOpen
  \bibfield  {author} {\bibinfo {author} {\bibfnamefont {P.}~\bibnamefont
  {Talkner}}\ and\ \bibinfo {author} {\bibfnamefont {P.}~\bibnamefont
  {H\"anggi}},\ }\href {\doibase 10.1103/PhysRevE.94.022143} {\bibfield
  {journal} {\bibinfo  {journal} {Phys. Rev. E}\ }\textbf {\bibinfo {volume}
  {94}},\ \bibinfo {pages} {022143} (\bibinfo {year} {2016})}\BibitemShut
  {NoStop}%
\bibitem [{\citenamefont {Talkner}\ and\ \citenamefont
  {H\"anggi}(2019)}]{TalknerHaenggiArXiv2019}%
  \BibitemOpen
  \bibfield  {author} {\bibinfo {author} {\bibfnamefont {P.}~\bibnamefont
  {Talkner}}\ and\ \bibinfo {author} {\bibfnamefont {P.}~\bibnamefont
  {H\"anggi}},\ }\href {https://arxiv.org/abs/1911.11660} {\bibfield  {journal}
  {\bibinfo  {journal} {arXiv: 1911.11660 (v2)}\ } (\bibinfo {year}
  {2019})}\BibitemShut {NoStop}%
\bibitem [{Note2()}]{Note2}%
  \BibitemOpen
  \bibinfo {note} {Towards the end of Sec.~IV.A.2 of Ref.~\cite
  {TalknerHaenggiArXiv2019}, T\&H also cite some literature using generalized
  measurements, but they still conclude that ``the measurements yet need to be
  error free''~\cite {TalknerHaenggiArXiv2019}.}\BibitemShut {Stop}%
\bibitem [{\citenamefont {Strasberg}\ and\ \citenamefont
  {Esposito}(2017)}]{StrasbergEspositoPRE2017}%
  \BibitemOpen
  \bibfield  {author} {\bibinfo {author} {\bibfnamefont {P.}~\bibnamefont
  {Strasberg}}\ and\ \bibinfo {author} {\bibfnamefont {M.}~\bibnamefont
  {Esposito}},\ }\href {\doibase 10.1103/PhysRevE.95.062101} {\bibfield
  {journal} {\bibinfo  {journal} {Phys. Rev. E}\ }\textbf {\bibinfo {volume}
  {95}},\ \bibinfo {pages} {062101} (\bibinfo {year} {2017})}\BibitemShut
  {NoStop}%
\bibitem [{\citenamefont {Kolmogorov}(2018)}]{KolmogorovBook2018}%
  \BibitemOpen
  \bibfield  {author} {\bibinfo {author} {\bibfnamefont {A.~N.}\ \bibnamefont
  {Kolmogorov}},\ }\href@noop {} {\emph {\bibinfo {title} {Foundations of the
  Theory of Probability}}},\ \bibinfo {edition} {second english edition}\ ed.\
  (\bibinfo  {publisher} {Dover Publications},\ \bibinfo {address} {Mineola,
  New York},\ \bibinfo {year} {2018})\BibitemShut {NoStop}%
\end{thebibliography}%
%\bibliography{/home/wiwi/Documents/references/books,/home/wiwi/Documents/references/open_systems,/home/wiwi/Documents/references/thermo,/home/wiwi/Documents/references/info_thermo,/home/wiwi/Documents/references/general_QM,/home/wiwi/Documents/references/math_phys,/home/wiwi/Documents/references/general_refs,/home/wiwi/Documents/references/equilibration}
%\bibliography{books,open_systems,thermo,general_QM,info_thermo,math_phys}

%%%%%%%%%%%%%%%%%%%%%%%%%%%%%%%%%%%%%%%%%%%%%%%%%%%%%%%%%%%%%%%%%%%%%%%%%%%%%%%%%%%%%%%%%%%%%%%%%%%%%%%%%%%%%%%%%%%%%%%%
%\appendix

\end{document}